\RequirePackage{fix-cm}
\documentclass[smallextended]{svjour3}       
\smartqed  
\usepackage{graphicx}
\newcommand{\arcsec}{$^{\prime\prime}$\,}
\newcommand{\arcmin}{$^{\prime}$\,}
\newcommand{\degr}{$^{\circ}$}
\newcommand{\sun}{_{\odot}}
\def\lesssim{\mathrel{\hbox{\rlap{\hbox{\lower3pt\hbox{$\sim$}}}\hbox{\raise2pt\hbox{$<$}}}}}
\def\farcs{\hbox{$.\!\!^{\prime\prime}$}}  
\def\day{\hbox{$^{\rm d}$}}                

\begin{document}
\sloppy

\title{Optimized Herschel/PACS photometer observing and
data reduction strategies for moving solar system targets$^{*}$
\thanks{{\it $^{*}$Herschel} is an ESA space observatory 
with science instruments provided by
European-led Principal Investigator consortia and with 
important participation from NASA.}}

\titlerunning{Herschel/PACS observing technics of TNOs}        

\author{Kiss, Cs. \and 
		M\"uller, Th.G. \and
		Vilenius, E. \and 
		P\'al, A. \and  
		Santos-Sanz, P. \and 
		Lellouch, E. \and
		Marton, G. \and
		Vereb\'elyi, E. \and
		Szalai, N. \and
		Hartogh, P. \and
		Stansberry, J. \and
		Henry, F. \and
		Delsanti, A.
}


\institute{Cs. Kiss$^{*}$, A. P\'al, G. Marton, E. Vereb\'elyi, N. Szalai \at
              Konkoly Observatory, MTA CSFK \\
              Tel.: +36-1-3919341\\
              Fax: +36-1-2754668\\
              \email{$^{*}$pkisscs@konkoly.hu}           
           \and
           Th.\,G. M\"uller and E. Vilenius \at
           Max-Planck-Institut f\"ur extraterrestrische Physik, Garching, Germany
           \and
           P. Santos-Sanz \at
           Instituto de Astrof\'\i sica de Andaluc{\'\i}a--CSIC, Granada, Spain
           \and
           E. Lellouch, F. Henry and A. Delsanti \at
           LESIA, Observatoire de Paris, Meudon, France
           \and
           P. Hartogh \at
           Max-Planck-Institut f\"ur Sonnensystemforschung, Katlenburg-Lindau, Germany
           \and
           J. Stansberry \at
           Space Telescope Science Institute, Baltimore, MD, USA
}

\date{Received: date / Accepted: date}

\maketitle

\begin{abstract}
The "TNOs are Cool!: A survey of the trans-Neptunian region" is a Herschel Open 
Time Key Program that aims to characterize planetary bodies
at the outskirts of the Solar System using PACS and SPIRE data, mostly taken as 
scan-maps. In this paper we summarize our PACS data reduction scheme
that uses a modified version of the
standard pipeline for basic data reduction, optimized for
faint, moving targets. 
Due to the low flux density of our targets the observations are
confusion noise limited or at least often affected by bright nearby
background sources at 100 and 160\,$\mu$m. 
To overcome these problems we developed techniques to characterize and 
eliminate the background at the positions of our targets and
a background matching technique to compensate
for pointing errors. We derive a variety of maps
as science data products that are used depending on the source flux and 
background levels and the scientific purpose.
Our techniques are also applicable to a wealth of other
Herschel solar system photometric observations, e.g. comets and 
near-Earth asteroids. The principles of our observing strategies and reduction
techniques for moving targets will also be applicable for similar surveys
of future infrared space projects.

\end{abstract}

\section{Introduction}

TNOs are frozen leftovers from the formation period of the outer 
Solar System. Due to their relatively small sizes and large distances 
little information can be earned from visual range observations only but
physical characteristics can be derived using visual range and thermal
emission observations together.
The cold surface of planetary bodies at the outer regions of the 
Solar System (20-50\,K) made the photometric instruments of the 
Herschel Space Observatory (Pilbratt et al., 2010) very well suited 
to survey these populations at far-infrared and submillimetre wavelenghts. 
The ``TNOs are Cool!'': A survey of the trans-Neptunian region Herschel Open 
Time Key Program (M\"uller et al., 2009) used the PACS 
(Poglitsch et al., 2010) and SPIRE (Griffin et al., 2010) instruments
to observe the thermal emission of trans-Neptunian objects and Centaurs. 
Due to the prominent sky background at these wavelengths, 
the observations are severely affected by confusion noise
at the longer PACS photometer bands (100 and 160\,$\mu$m)
and the SPIRE bands (250, 350 and 500\,$\mu$m), where the resolving power
is poorer. However, the relatively fast apparent motion of our targets 
(typically a few arcsecs per hour) made it possible to develop 
observation schemes and data reduction techniques that could 
fully exploit the potential of multi-epoch observations and 
eliminate the background (and hence the confusion noise) very
effectively.
Due to the high sensitivity of PACS in this programme --
in agreement with our original goals -- 
we were able to sample about 10\% of the known TNOs and determine
their main physical characteristics (size, albedo, surface properties) 
with the synergy of ground-based optical and space-born far-infrared data.
The main science goals of our program were
(i) simultaneous measurement of size and albedo of a large sample of targets;
(ii) Determination of densities of binary TNOs based on measured effective sizes;
(iii) Constraining of thermal and surface properties;
(iv) Measurement of thermal lightcurves of few objects by continuously observing 
them throughout an entire rotational period. 
Our science results are published in a series of about
 15 publications, including recent ones by Santos-Sanz et al. (2012),
 Mommert et al. (2012), Vilenius et al. (2012), Pal et al. (2012),
 Kiss et al. (2013), Fornasier et al. (2013), Lellouch et al. (2013),
 and two recently submitted papers by Duffard et al. (2013), and
 Vilenius et al. (2013).
The large amount and high complexity of our 
data required the development of our own pipeline processing, and 
some special data reduction 
techniques were developed to fully exploit the capabilities of 
our multi-epoch observations. 
In this paper, in addition to the description of our original observation 
planning strategies, we summarize the main steps of our PACS scan-map data
reduction and the validation of the processing scheme used in our programme.
We also describe our data products that are derived from the raw science data and 
that are used to obtain the final flux densities. 

\section{Observations}

\subsection{Open Time Key Program and related observations}

The observations of the {\it Herschel} key programme
``TNOs are Cool: A survey of the trans-Neptunian region'' 
Herschel Open Time Key Program (M\"uller et al., 2009) 
consist of Science Demonstration Phase 
(SDP) and a Routine Science Phase observations, 
with a total time allocation of 403.3 hours (including all overheads), 
of which 30.6 hours were used in the SDP between November 2009 and January 2010,
and all observations were performed before by October 2012. 
All SDP targets were observed again during the routine phase.
We used 95\% of the total time for photometry observations with PACS (60-210\,$\mu$m) 
and the rest with SPIRE (200-670$\mu$m). 
The data reduction of our SPIRE observations is presented in Fornasier et al. (2013).

In addition to the OTKP measurements, there were several successful
Open Time Programs and Director's Discretionary Time (DDT) observations  
related to the ``TNOs are Cool!'' OTKP. For example, 
Eris and Quaoar thermal light curves were obtained in the open time program
"Probing the extremes of the outer Solar System: short-term variability of the 
largest, the densest and the most distant TNOs from PACS photometry"
(PI: E. Vilenius; see Kiss et al., 2012), 
we investigated "Pluto's seasonal evolution and surface 
thermal properties" (PI: E. Lellouch), 
and observed two objects that move on peculiar orbits, 2012\,DR$_{30}$ and
2013\,AZ$_{60}$ in two dedicated DDTs in May 2012 and 
April 2013, respectively (see Kiss et al., 2013 for the 2012\,DR$_{30}$
results). All these observations used the same reduction pipeline to
obtain the final flux densities of the targets. More details 
on these additional observations are presented in the respective papers. 
A summary of all ``TNOs are Cool!'' Open Time Key Program observations
can be found at the following webpage:

\medskip

{\large\it http://kisag.konkoly.hu/tnodatareductionsummary}

\subsection{Observation design}

We used the Standard Thermal Model (Lebofsky et al., 1986, and references therein) 
to predict flux densities of our targets
in the PACS bands. Based on earlier \emph{Spitzer} work 
(Stansberry et al., 2008) we adopted
a geometric albedo of 0.08 for those targets which did not have previous \emph{Spitzer}
results, and used a hybrid-STM, in which the beaming parameter differs from the canonical
STM beaming parameter. We used $\eta=1.25$ 
(Stansberry et al., 2008) for observation planning purposes.
The predicted thermal fluxes depend on the sizes, which are connected to the 
assumed geometric albedo and the absolute V-magnitudes via
\begin{equation}
D=\frac{2a}{\sqrt{p_V}} \times 10^{\frac{1}{5} \left( m_\mathrm{\sun} - H_V \right)},
\label{pA}
\end{equation}
where $D$ is the area equivalent diameter of the TNO assumed to be spherical, 
$a$ is the distance of one astronomical unit, $p_V$ is the \emph{assumed} 
geometric albedo, $m_\mathrm{\sun}$ is the apparent V-magnitude of the Sun, 
and $H_V$ is the absolute V-magnitude of the TNO.

The submitted and accepted OTKP target list consisted of 137 TNOs and 
Centaurs, and in addition the giant planet moons Phoebe and Sycorax. We have observed 132
of them, including the two moons. The main reasons for not observing seven 
targets were too uncertain astrometry and too low predicted flux.
The absolute V-magnitudes ($H_{\mathrm{V}}$) used in the planning of our sample 
were $<10.8$\, mag 
($<8.2$\, mag if Centaurs are excluded). 
In the scientific analysis of observed targets we use the latest optical photometry 
available in the literature or in data bases, or determine absolute visual magnitudes
based on data from these sources. For some targets, there has been a significant 
change in $H_{\mathrm{V}}$ compared to the estimate used during the planning.

The observation duration and timing constraints have been planned individually for 
each target. 
All of the targets observed had predicted astrometric $3\sigma$ 
uncertainties 
less than 10\arcsec at the time of the \emph{Herschel} observations
(David Trilling, {\it priv.~comm.}). In the routine phase we used the AstDys 
web service\footnote{Asteroids Dynamic Site by A. Milani, Z. Knezevic, O. Arratia 
et al., http://hamilton.dm.unipi.it/astdys/, calculations based 
on the OrbFit software}. 


\subsection{Observing modes \label{sect:obsmodes}}

We specified the astronomical observation requests (AOR) 
in HSpot, a tool provided by the Herschel Science Ground Segment 
Consortium. In the PACS photometer AORs a selection is made to observe
either the blue+red or the green+red channels (the red channel data are taken
simultaneously whichever short wavelength filter is chosen). 
The sensitivity of the blue
channel is usually limited by instrument noise, while the red channel
is confusion-noise limited (Poglitsch et al., 2010). 
The sensitivity in the green channel can
be dominated by either of them, depending on the depth and the region of the sky
of the observation. We optimized the timing of the observations 
by selecting the visibility window in which the far-infrared confusion 
noise (Kiss et al., 2005) was the lowest in the green channel. 

Since all of our targets, including complete multiple TNO systems, 
have apparent sizes smaller than the PACS spatial resolution, 
we aimed to measure the disk-integrated flux density using
either the point-source mode (usually referred to as "chop-nod" mode) 
or the mini-scan map mode. Both options were tested extensively 
during the science demonstration phase (SDP) of the Herschel mission. 

\paragraph{Chop-nod mode:}

Details on the chop-nod mode can be found in the PACS Observer's 
Manual\footnote{http://herschel.esac.esa.int/Docs/PACS/html/pacs\_om.html} and 
in the observing mode release 
note\footnote{http://herschel.esac.esa.int/twiki/pub/Public/PacsCalibrationWeb/PhotMiniScan\_ReleaseNote\_20101112.pdf}. 
In this mode the target is moved to different locations on the detector 
array by a chopper mirror as well as by nodding the telescope pointing. 
The source remains on the array all the time. 
One of the few parameters left for the astronomer to choose, in addition to
the number of repetitions, is whether dithering is used.
We used the PS mode (10, 16 or 36 repetitions corresponding to durations of 0.4, 
0.7 or 1.6 hours) with dithering
only for six targets during the SDP (M\"uller et al., 2010\footnote{Targets: 
42355 Typhon (two observations), 79360 Sila,
82075 (2000 YW$_{134}$), 126154 (2001 YH$_{140}$) and
208996 (2003 AZ$_{84}$).}, Lim et al., 2010\footnote{Target 136472 Makemake 
(two observations).}). 
The flux densities of our targets observed in this mode were 
derived with the standard chop-nod reduction pipeline script 
available in HIPE (Ott 2010), using the latest available version. 
The final chop-nod mode maps contain images of the target, as
 well as images of all background sources 
 in the neighbouring field, forming a specific structure. 
This makes it very challenging to 
identify the target and perform suitable photometry in most cases,
especially for faint sources close to the confusion limit.   
As the usability of this mode was very much restricted according to 
our tests in the SDP,
and as the mini scan-maps showed a much better overall efficiency
(as discussed in detail below), 
at the end of the SDP we decided to switch to the 
scan-map mode as our default observing mode. 
Note that all of our chop-nod SDP targets were later re-observed in the routine phase
in the mini-scan map mode. The chop-nod measurements were reduced using the
latest version of the standard pipeline, and no
further combined products were derived from the Level-2 chop-nod maps.
Since the chop-nod mode is well calibrated, most recently by Nielbock et al. (this issue), 
chop-nod data can be used in radiometric modeling techniques in combination 
with the mini-scan map data taken at a different epoch. Science results 
using chop-nod data of the "TNOs are Cool!" Open Time Key Program have been
published in M\"uller et al. (2010) and Lim et al. (2010).

\paragraph{mini scan-maps:}

\begin{figure*}
\hbox{ \includegraphics[height=47mm]{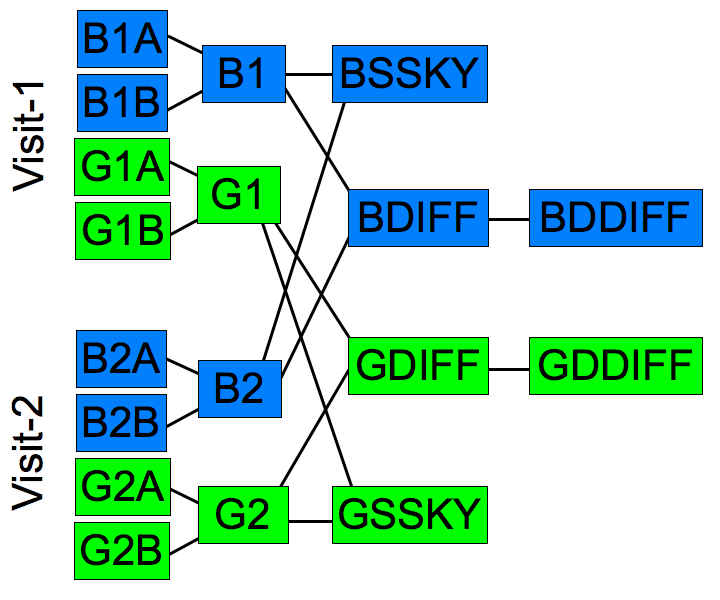}
       \includegraphics[height=47mm]{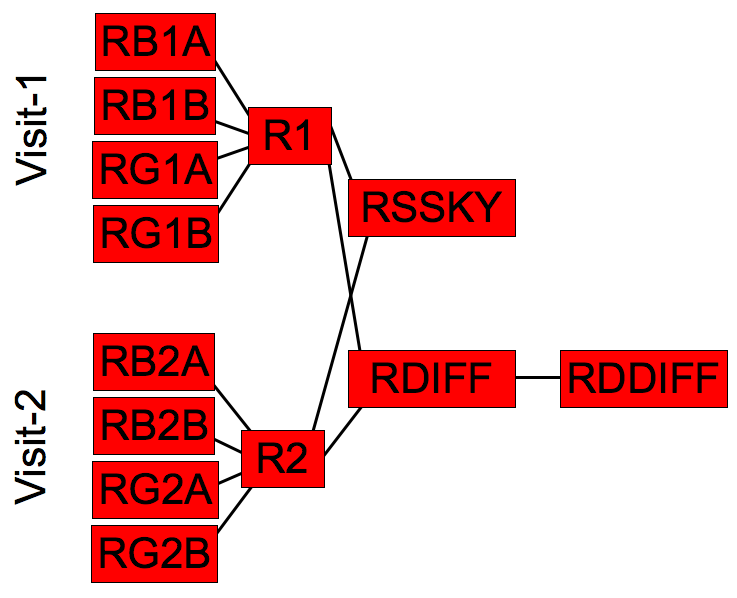}}
\caption{Outline of our standard observing and image derivation scheme. 
The single maps (first column)
are combined to obtain the co-added, single visit maps (second column), and these
co-added maps are used to produce the different science data products 
(SSKY, DIFF and DDIFF maps) that are used to obtain the final fluxes. The left and
right panels of the figure show the scheme separately for the short wavelength
(70/100\,$\mu$m, or blue/green) and for the long wavelength (160\,$\mu$m or red)
PACS channels. In each box in the first letter marks the filter (B=blue, G=green, R=red),
the second marks the epoch (1\,=\,Visit-1, 2\,=\,Visit-2), the third 
marks the scan direction (A\,=\,70\,deg, B\,=\,110\,deg). 
In the case of the red filter sequences the
double letters (RB or RG) marks the corresponding short/long wavelength 
filter combination. The SSKY, DIFF and DDIFF labels correspond to the
supersky-subtracted, differential and double-differential images, respectively. }
\label{fig:scheme}       
\end{figure*}

During the SDP we tested the mini scan-map 
mode\footnote{http://herschel.esac.esa.int/twiki/pub/Public/PacsCalibrationWeb/\\
PhotMiniScan\_ReleaseNote\_20101112.pdf} with two small
maps per target, observed at different scan angles with respect to the detector array.
The SDP light curves observations of (136108) Haumea were also performed in
this mode (Lellouch et al. 2010). 
This mini-scan flavor of the scan-map mode also became the recommended mode 
for faint point sources (Poglitsch et al., 2010) due to its better
overall efficiency in the case of relatively faint targets, such as ours. 
In this mode the pointing of the telescope is slewed
at a constant speed over parallel lines, or ``legs''. 
We used 10 scan legs in each AOR, separated by 4\arcsec. The length of each leg was 
3.0\arcmin except during the SDP and in the beginning
of the routine phase, when 3.5\arcmin and 2.5\arcmin were used. The selected 
slewing speed was 20\arcsec\,s$^{-1}$ except for 22 AORs in January 2010 when 
the fast scan speed of 60\arcsec\,$s^{-1}$ was tested.
During the SDP different observing strategies were in use but
in the routine phase we mostly used a constant sequence of AORs 
(except for lightcurve observations). For a given channel selection (blue or green)
we grouped pairs of AORs, with scan orientations of 70\degr and 110\degr with respect
to the detector array (Scan-A and Scan-B), 
in order to make optimal use of the rectangular shape of the 
detector. Thus, during a single visit of a target we grouped 4 AORs to be observed 
in sequence: two AORs in different scan directions with the same 
short wavelength PACS filter, and then this sequence repeated for the 
second short wavelength channel. The outline of this scheme is 
presented in Fig.~\ref{fig:scheme}. 

Within each AOR the observation of the mini-map was repeated from one to six times,
depending on the predicted flux of the target, as described above.
Each target was visited twice with similar AORs repeated in both visits.
The timing of the visits (Visit-1 and Visit-2) was such
that the target had moved 30--50\arcsec between the visits so that the target position
during the second visit is within the high-coverage area of the map from the first visit.
This allows us to determine the background for the two source positions,
as the Visit-1 and Visit-2 images serve as mutual backgrounds.
The typical time difference between Visit-1 and Visit-2 was in the
order of a few hours for Centaurs and 1--1.5\,days for TNOs due to their
different apparent velocities. In the case of fast moving near-Earth objects
we could use individual repetitions within the same measurement as mutual 
backgrounds in two or more visits, as was done e.g. in the case 
of (162173) 1999\,JU$_3$ (O'Rourke et al., 2012).  


\section{Data reduction of mini scan-maps}
\label{sect:datareduction}



We use a modified version of the PACS pipeline for basic data reduction 
of scan-maps\footnote{For details, see the PACS Data Reduction Guide:
http://herschel.esac.esa.int/twiki/bin/viewfile/Public/PacsCalibrationWeb?rev=1;filename=PDRG\_Dec2011.pdf}
 (producing single images per OBSID) 
from raw data to Level-2 maps (for the definition of the 
Herschel/PACS data product levels, see the PACS Observer's Manual). 
We applied the following main parameters
in HIPE (for a summary of the PACS photometer scan-maps 
calibration, see Balog et al., this issue):
\begin{itemize}
\item Slews are selected on scan speed, usually between 15 and 25\arcsec\,$s^{-1}$ 
(our maps are predominantly observed with 20\arcsec\,$s^{-1}$ scan speed). 
\item High-pass filter width of 8, 9 and 16 are used 
at 70, 100 and 160\,$\mu$m, respectively 
(high pass filter width sets the number of frames [2n+1] used
for median subtraction from the detector timeline; see 
Popesso et al., 2012 and Balog et al., 2013, this issue, 
for a detailed description of the method)
\item Masking pixels above 2-sigma, and at the source position with 2xFWHM radius
\item We apply second level deglitching with nsigma=30, the
sigma-clipping parameter of this deglitching method working on the map level
(see the PACS Data Reduction Guide for more details). 
At the same time, the {\it multiresolution median transform} 
(MMT) deglitching task is disabled (see Vavrek et al., 2008). 
\item Despite that we observe solar system targets, we {\it do not correct} for 
the apparent motion of target to obtain the data products described below. 
In most cases the apparent displacement of our target 
during a single visit is notably smaller than the blue 
(shortest wavelength) FWHM, and such a correction would 
degrade the  performance of our background elimination 
techniques at the highest spatial frequencies. 
For those targets that have a displacement larger than 
0.5{\arcsec} per individual visit, we use the standard, 
proper motion corrected reduction scheme. 
\end{itemize}

We apply the drizzle method to project the time-line data and produce the
single maps using the {\it photProject()} task in HIPE, with
a pixel fraction parameter of 1.0 in most cases. 
We use default pixel sizes of 1\farcs1, 1\farcs4 and 2\farcs1 in the
PACS 70, 100 and 160\,$\mu$m bands, respectively (in some special cases different 
pixel sizes or pixel fraction parameters were also used).

The data reduction is performed on dedicated computers which have the 
sufficient amount of memory (up to 128\,GB) and the necessary CPU performance. 
These computers are located at Max-Planck-Insititut f\"ur Sonnensystemforshung 
(Katlenburg-Lindau, Germany) and Konkoly Observatory (Budapest, Hungary). 

\section{"TNOs are Cool!" image products}

As a further step in our data reduction, we combine the single 
maps obtained in Visit-1 and Visit-2 with the aim to reduce the effect 
of the background as much as possible. We produce the following image products:
\begin{itemize}
\item Co-added images (from the Scan-A and Scan-B images of the same, single visit)
\item Differential images (from the co-added images, DIFF). 
Optimal coordinate offsets are determined with the "background matching" method
\item Super-sky subtracted images (from the co-added images, SSKY)
\item Double differential (DDIFF) images, created from the differential images, 
using "source matching" to determine the ideal offsets
\end{itemize}

We are summarizing the details of these products and the necessary intermediate 
steps {below}.  

\paragraph{Co-added images:}

Co-added images are generated using the maps of the individual OBSIDs in
a specific band and in a single visit. In the case of both the 
blue and the green band we co-add two maps, the Scan-A and Scan-B images
(B1\,=\,B1A+B1B, G1\,=\,G1A+G1B, etc., according to the 
scheme presented in Fig.~\ref{fig:scheme}). 
In the red band, all the four red maps (taken in parallel with 
blue/green and scan/cross-scan) are co-added (R1\,=\,RB1A+RB1B+RG1A+RG1B, etc.). 
This processing step is performed in 
IDL\footnote{Interactive Data Language, Research Systems Inc.} 
considering the coverage values of each pixel as weights. 
This, in principle, is identical with the co-added images obtained using 
the MosaicTask() function in HIPE. These co-added images are the bases 
of the further processing steps and data products. 

\paragraph{Supersky-subtracted images:}

\begin{figure*}
  \includegraphics[width=0.99\textwidth]{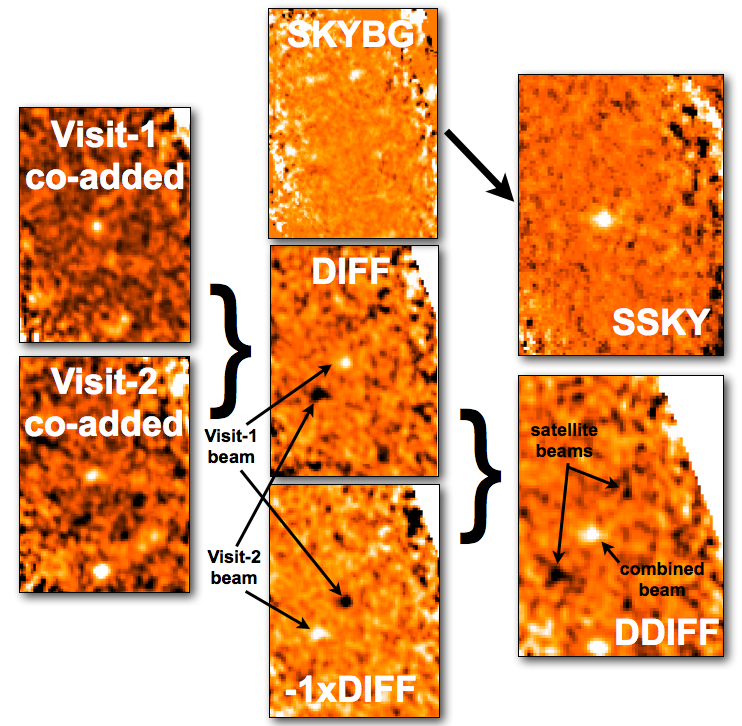}
\caption{{Demonstration of mini-scan map image processing steps and
products from the co-added image level with the 160\,$\mu$m images of 
the Centaur 2002\,GZ$_{32}$, a relatively faint target in this band
(see also Fig.~1 and Sect.~4 for the detailed data reduction scheme)}}
\label{fig:ss}       
\end{figure*}

To create the so-called supersky-subtracted images first a background map is 
generated using the single maps. To do this we “mask” the target in each single map and
co-add the maps in the sky coordinate system. This step produces a background map 
without the target. The background map is subtracted from the single maps 
producing background-subtracted single maps. Finally the background-subtracted maps 
are co-added in the target frame, producing the final
combined map on which photometry is performed (this method has previously 
been described in detail and demonstrated with sample images in Santos-Sanz et al., 
2012). A feature of this method is that at the masked locations the 
signal-to-noise ratio is lower than at the other parts of the image, since 
only the data of a single visit can be used here. An example 
is presented in Fig.~\ref{fig:ss} for the Centaur 2002\,GZ$_{32}$. 
This background subtraction technique was originally developed for the 
Spitzer/MIPS observations of trans-Neptunian objects (Stansberry et al. 2008),
and we kept the original scheme to obtain the final images, but applied
the background matching method to correct for coordinate frames 
offsets between the two visits (see below). 

\paragraph{Differential images and background matching:}

\begin{figure*}
\hbox{
  \includegraphics[width=0.62\textwidth]{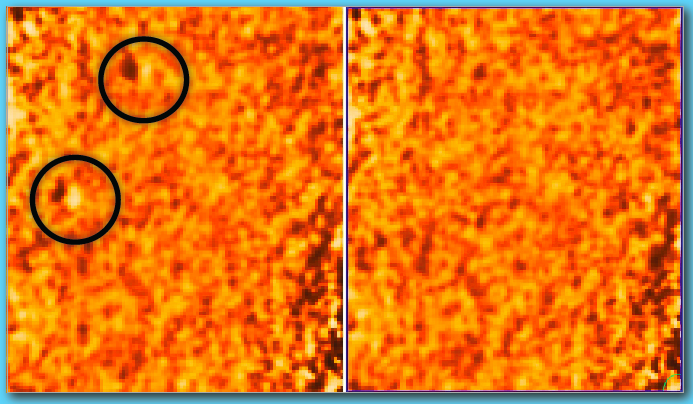}
  \includegraphics[width=0.38\textwidth]{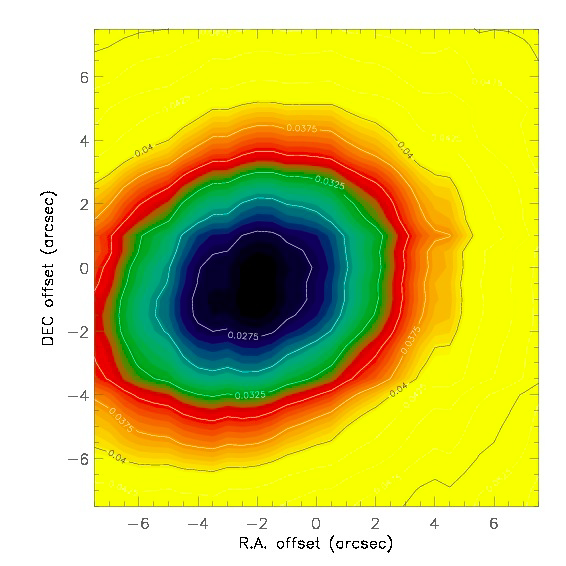}}
\caption{Left panel: Original (left) and “background matched” (right)
differential images of the same pair of 70\,$\mu$m images. 
Right panel: Contour map of residual noise as a function of coordinate offsets. 
The optimal offsets were -2\farcs5 and -1\farcs0 in R.A. and DEC, 
respectively.}
\label{fig:bgmatch}       
\end{figure*}

Background matching is used to correct for the small offsets in the coordinate 
frames of the Visit-1 and Visit-2 images when obtaining the differential
 image, which is simply the difference of the combined Visit-1 and Visit-2
 images in the respective bands (BDIFF\,=\,B1--B2, etc., see also Fig.~1).   
 Incorrect offsets can easily be identified by the appearance of 
 positive/negative spot pairs (see the left panel of Fig.~\ref{fig:bgmatch} 
 above, marked with black ovals) -- these are completely eliminated on the
corrected image (right panel of the same image). 
The offset to be applied can be determined using images of systematically shifted
coordinate frames and then determining the offset which provides the smallest standard
deviation of flux values in a pre-defined coverage interval 
(typically 0.3\,$<$\,coverage\,$<$\,0.9, see the contour map of Fig.~\ref{fig:bgmatch}).
Our tests have proved that the same offset is obtained using any of the three PACS bands, 
however, in most cases the offset can be most readily determined using 
the 160\,$\mu$m images, due to the strong sky background w.r.t. the instrument noise.

\begin{figure*}
  \includegraphics[width=0.99\textwidth]{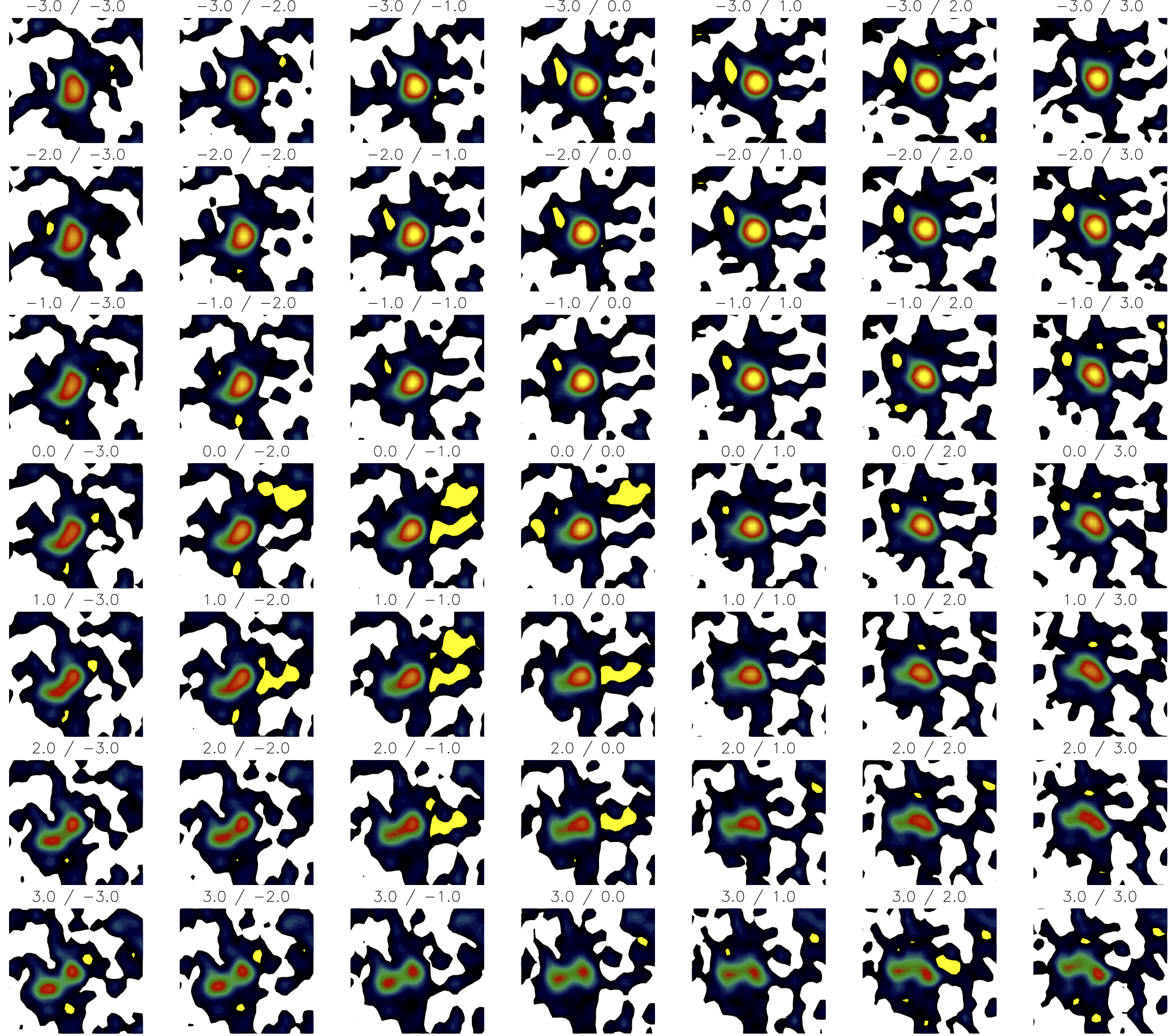}
\caption{Source matching for Ixion (green band) 
to determine the optimal offset for the final DDIFF image. Wrong offsets
can easily be identified by the distorted or double-peaked shape of the 
combined image. The numbers above the stamp figures correspond to the
actual offsets in {\it arcsec} units (the optimal offsets are
--2\farcs0 and +1\farcs0 in this case).}
\label{fig:sourcematching}       
\end{figure*}

\paragraph{Double-differential images:}

A double-differential (DDIFF) image is made of the DIFF image of a target at a 
specific wavelength. The disadvantage of the DIFF image is that the 
images of the target appears as two separated beams (one positive and one
negative), corresponding to the two visits. 
To produce a DDIFF image, first the DIFF image is "folded" (multiplied by $-1$). 
The folded image is shifted in a way that the location of the two positive beams 
of the target match on the original and the folded image (see Fig.~\ref{fig:ss}).
Then, the original and the folded/shifted DIFF images are co-added:
\begin{equation}
DDIFF(\underline{x})\,=\,DIFF(\underline{x})
-DIFF(\underline{x}+\underline{\theta})
\end{equation}
where the optimal offset $\underline{\theta}$ is determined with the
source matching method (see below). 
The DDIFF image contains a positive beam with the {\it total flux} 
of the target and two negative beams at the sides with "half" of the total flux
\footnote{well, this is not exactly half. The two negative beams contain the half flux
of the first and second visits separately, which is half the co-added DDIFF
flux only if the flux of the target has not changed between the two visits}. 
It is a clear advantage of this method that the photometry can be performed on
a single beam, and one does not have to combine the flux of two beams as in the
case of the DIFF images. 
In the case of the DDIFF images the noise is increased by a factor of $\sqrt{2}$ 
when compared to the corresponding DIFF image, and flux variations between the two
visits are flattened out. However, the signal-to-noise of the target is improved by 
$\sqrt{2}$ with respect to a single DIFF image which is very important in 
detecting faint targets. This method has proved to provide the best performance in 
the detection of very faint sources ($<$\,2\,mJy at 70\,$\mu$m),
superior to the DIFF or SSKY images. 


\paragraph{Source matching:}
Background matching (see above) provided offsets for coordinate frame differences 
in the two visits, but positional differences may still remain due to e.g. not 
well known positions of the target, and wrong offsets lead to distorted shapes of 
the target image when the images of the two visits are combined 
to obtain double-differential images. 
"Source matching" determines the optimal offset ($\underline{\theta}$) 
that the original and folded 
DIFF images have to be shifted with to obtain the best matching of the centroids
of the targets when we combine them to produce the 
DDIFF images. Typical offsets are a few arcseconds, 
we use the $\pm$4\arcsec range both in R.A. and DEC to determine the offset.
We demonstrate the method in Fig.~\ref{fig:sourcematching} for Ixion.  
For relatively bright targets (a few tens of mJy) the source
matching correction typically increase the flux by $\lesssim$\,10 per cent compared
to the uncorrected case -- in these cases the optimal offsets are in the 
order of 2\arcsec. However, to detect very faint targets, source matching 
is a necessary step to detect the target at all.

\medskip

The source matching optimized double-differential method
has proved to provide the best performance in 
the detection of very faint sources 
($\lesssim$\,2\,mJy at 70 and 100\,$\mu$m and $\lesssim$\,5\,mJy 
at 160\,$\mu$m),
superior to the DIFF or SSKY images. The clear difference between the
DIFF and DDIFF images is the $\sqrt{2}$ signal-to-noise improvement.
Similarly, in the case of SSKY images the noise level is higher at 
the masked area in the vicinity of the target, as only the background 
information of a single visit is used. For moderately bright targets
(at least a few mJy flux density) both the DDIFF and SSKY products
are used to extract fluxes.



An automatic derivation of the combined Open Time Key Program
data products is performed using the {\it FITSH} package (P\'al, 2012). 

\section{Photometry}
\label{sect:photom}

\paragraph{Aperture photometry} of our targets can be
performed using various tools (IDL/DAOPHOT, IRAF, HIPE) that provide 
practically identical results, as it has been tested at the early 
phases of our programme. As the default image pixel size 
is adjusted to the actual FWHM in all PACS bands 
(1\farcs1, 1\farcs4 and 2\farcs1 in the blue, green and red bands, 
respectively), an ideal aperture of 4-5 pixels in radius has been
identified in all bands using flux growth curves of several targets. 
The encircled energy fraction coefficients -- that correct
for the flux outside the measuring aperture -- are taken from the
PACS Observer's Manual. 

\paragraph{Absolute photometric accuracy}
Although the final, level-2.0 images after the pipeline processing
are already flux calibrated, it is important to check whether our 
processing introduces any discrepancies with respect to the 
official flux calibration. It is especially important in our case 
since the flux calibration of the PACS photometer is based on a few 
bright stars and asteroids, and no faint star (in the order of 
$\sim$\,100\,mJy at 70\,$\mu$m or below) 
is used officially as a calibrator. Our targets are significantly fainter 
than the calibrator objects, and in principle they may require a different 
flux calibration. We selected a few faint star calibration measurements
with available photospheric flux predictions for the PACS 70, 100 and
160\,$\mu$m bands (HD\,139669: Shirahat et al., 2009; 
$\gamma$\,Dra, HD\,170693 and HD\,138265: Gordon et al., 2009). 
The measurement were taken  
from the Herschel Archive and processed in the same way 
as would be done with the TNO measurements. 
After the level-2.0 products (maps) are created, we
performed aperture photometry again on the central source 
in the same way as in the case of the TNO measurements. We characterise the
absolute calibration accuracy by two numbers: a coefficient 
that the measured flux has to be multiplied with to obtain the predicted flux
(F$_{pred}$\,=\,r$_{cal}\times$F$_{meas}$) 
and the relative uncertainty of the measured to predicted fluxes
($\sigma_{cal}$). The r$_{cal}$ values we obtained are 1.03, 1.01 and 0.98 for
blue, green and red bands, respectively, while the $\sigma_{cal}$
values are 0.9\%, 1.5\% and 5.6\%, in very good agreement with the 
generaly quoted 5\% absolute accuracy of the PACS photometer flux 
calibration. 

The absolute photometric accuracy of our full processing 
scheme -- that includes differential, double differential 
or supersky-subtracted images -- cannot be tested entirely using standard star 
measurements, as these objects do not move with respect to the background, 
making these types of images meaningless to produce. 
As the correctness of the flux calibration was approved in the first step
(see above), in a second step we used the scan-map measurements of some bright TNOs 
and Centaurs (Quaoar, Orcus, Ixion and Chariklo, among others) 
to test whether flux densities are 
kept during the generation of our combined data products from the differential 
images to the double differential or supersky-subtracted images. The requirement 
was that, as the background is negligible for these bright targets, the average 
flux densities obtained at the original, single visit images must be the same as 
that of the combined products within the photometric uncertainties. According to 
these tests the combined data products fulfil this requirement
(relative accuracies are better then 5\% in all PACS bands). 

\paragraph{Photometric uncertainty} is determined using the 
"implanted source" method in the case of all data products. In this method 
we place 200 artificial sources on the image 
(a single one at a time) and this artificial source has a spatial flux distribution
shape of the PACS PSF in the actual band. 
Then the same type of aperture photometry is performed on each implanted source
as on the target. The sources are placed in regions with coverage values within a 
fixed interval (typically 0.3\,$<$\,coverage\,$<$\,0.9) -- this excludes
the vicinity of the target as well as the edges of the image where the 
coverage is low and therefore the noise is high.  
The photometric uncertainty of the map is taken as the standard deviation of the 
distribution of the artificial, implanted source fluxes (the distribution of 
these fluxes turned out to be very close to Gaussian in all cases).

\section{Detection statistics}
\label{sect:detectionstatistics}
In the ``TNOs are Cool!": A Survey of the trans-Neptunian region
Herschel Open Time Key Program we observed 132 targets, in total 
1131 observations (1089 in "KPOT\_thmuelle\_1", 
2 in "AOTVAL\_thmuelle\_2" and 40 in "SDP\_thmuelle\_3"). 
A good majority of the targets ($>$90\%) are detected in at least one band
(in the blue band in almost all cases), 
about 50\% of them are detected in all the three PACS bands.
With our techniques we managed to reach 0.6, 0.9 and 1.6\,mJy flux 
uncertainties using the combined products (DDIFF images) of 5-repetition single 
maps in the 70, 100 and 160\,$\mu$m PACS bands, respectively. 


\section{Summary}

In this paper we have demonstrated that the observation planning, data reduction pipeline 
and the related combined data products that we use in our open time 
key program are very effective tools to observe faint, moving solar system targets.  
%
%
Our methods, including observation planning, target selection,
observing template setup, data reduction and product generation, could also
be used to observe moving Solar System targets in future infrared space programs, 
like the SPICA mission (Nakagawa et al., 2012), to be launched in 2020.  
The applicability of these techniques for SPICA observations is summarized 
in Kiss et al. (2013b). More information on our open time key 
program can be found at the following webpage:

\medskip

{\large\it http://www.mpe.mpg.de/$\sim$tmueller/tno\_public/index.htm}

\begin{acknowledgements}
This work has been supported by the Hungarian Research Fund (OTKA) 
grant K~104607, the PECS-98073 grant of the European Space Agency (ESA) and the 
Hungarian Space Office and the Bolyai Research Fellowship of the 
Hungarian Academy of Sciences. E.V. acknowledges the support of the German DLR project 
\#\,50\,OR\,1108. The work of A.P. has been supported by the “Lend\"ulet” grant 
\#\,LP2012-31/2012 of the Hungarian Academy of Sciences. 
We are indebted to the Herschel observation planning team for their enthusiastic 
work during the active phase of the program which has highly 
contributed to the great success of the observations. 
We also thank the referee for providing constructive comments 
and help in improving this paper.

\end{acknowledgements}



\end{document}